\documentstyle{aipproc}
\begin{document}
\title{The Experiment Road to the Heavier Quarks
and Other Heavy Objects}

\author{Jeffrey A. Appel}
\address{Fermilab, PO Box 500, Batavia, IL 60510, USA\\
E-mail: appel@fnal.gov}

\maketitle

\input epsf

\begin{abstract}
After a brief history of heavy quarks, I will discuss charm, bottom, and
top quarks in turn.  For each one, I discuss its first observation, and
then what we have learned about production, hadronization, and decays -
and what these have taught us about the underlying physics.  I will also
point out remaining open issues.  For this series of lectures, the charm
quark will be emphasized.  It is the first of the heavy quarks, and its
study is where many of the techniques and issues first appeared. Only very
brief mention is made of CP violation in the bottom-quark system since that
topic is the subject of a separate series of lectures by Gabriel
Lopez.  As the three quarks are reviewed, a pattern of techniques and
lessons emerges.  These are identified, and then briefly considered in
the context of anticipated physics signals of the future; e.g., for
Higgs and SUSY particles.
\end{abstract}

\section*{A Brief History of Heavy Quarks}

\subsection*{Today's Elementary Particles}

Today's picture of the most elementary particles is composed of
six quark types, six lepton types, and four force carriers.  The
quarks and leptons come in three generations, each generation with a
pair of quarks (one with charge + 2/3, and one with charge - 1/3) and a
pair of leptons (one charged, and one neutral).  See Table
\ref{particles}.  The quarks and leptons have spin 1/2, and couple
variously to the spin 1 force carriers: gluons, photons, and charged
and neutral weak bosons ($W^+$, $W^-$, and $Z$).  Unlike the leptons,
the quarks appear to come in three varieties, called colors.
Perhaps the number of colors is related to the quarks' third-integer
charges.  Also, quarks never appear in isolation, being permanently
confined, for example, in meson and baryon combinations (quark-antiquark
and three-quark combinations, respectively).

\begin{table}
\caption{Today's elementary particles, by type vs generation.}
\label{particles}
\begin{tabular}{|c|c|c|c|c|}
\hline
Type  & charge  &$1^{st}$ Generation &$2^{nd}$ Generation & $3^{rd}$
Generation \\
\hline\hline
up-type   & + 2/3 & ${\it u}$ & ${\it c}$ & ${\it t}$   \\
quarks    &       &    up     &   charm   &     top     \\
\hline
down-type & - 1/3 & ${\it d}$ & ${\it s}$ & ${\it b}$   \\     
quarks    &       &  down     &  strange  &   bottom    \\
\hline\hline
neutral   &  0    & ${\nu_e}$ &${\nu_{\mu}}$ & ${\nu_{\tau}}$   \\     
leptons   &       & ${\it e}$ neutrino & $\mu$ neutrino & $\tau$ 
neutrino \\
\hline
charged   &  -1   & ${\it e}$ & ${\mu}$   & ${\tau}$    \\     
leptons   &       & electron  &   muon    &    tau      \\
\hline
\end{tabular}
\end{table}

These quark, lepton, and force-carrying particles are the foundation of
the so-called Standard Model of particle physics.  It is widely and
completely accepted by the community.  However, it was not always that
way.  The acceptance of the quark picture of elementary particles owes a
great deal to the discovery and understanding of the heavier quarks.
Quarks were unexpected, even not accepted when I was a student.  The
idea of ``partons'' obeying SU(3) symmetry~\cite{gellmann1,bj,nambu}
was a mathematical tool at best, not a physical
reality!~\cite{gellmann2}
There were alternate possibilities for underlying structure; e.g.,
based on shapes, on how things are put together, and on the
``bootstrap'' model.~\cite{chew}
Perhaps the hadrons we see in the laboratory are each made of
combinations of all the others, with no special subset being the most
elementary.

The first confirmation of the idea of quarks came after the prediction
by Murray Gell-Mann~\cite{gellmann2} of what we call the omega minus,
understood now to be a baryon made of three strange quarks.  This
otherwise unheralded particle, was discovered in a 1964 Brookhaven
bubble-chamber experiment headed by Nick Samios.~\cite{samios}

\subsection*{The Revolution of November, 1974}

The real watershed in thinking began with the announcement in November,
1974, that teams of physicists had observed a rather narrow resonance at 
a mass of 3.1 GeV/$c^2$.   The resonance was seen in both
hadroproduction~\cite{ting} and $e^+e^-$ annihilation.~\cite{richter}
The resonance still carries the dual name $J/\psi$ from these concurrent
observations.  The quickly-accepted model explaining this narrow
resonance was that it is made up of a new quark-antiquark pair.  These
new quarks were characterized by a new quantum number, called ``charm.''
The existence of a new quark implied a whole spectrum of new particles
containing at least one charm quark.  Examples of these were soon
discovered in $e^+e^-$ collisions by the Mark I Collaboration at 
SLAC.~\cite{D0Dplus}

\subsection*{The Growing Variety of Quarks}

With the discovery of charm particles, there was a feeling that the
quark picture of matter was complete, perhaps like the feeling at
the beginning of the century when the atomic nature of matter was first
understood.  Nevertheless, only three years after the discovery of
charm, an even heavier resonance was observed in proton-nucleus
collisions at Fermilab.~\cite{discoveryY}  This resonance, at a mass of  
about 9.5 GeV/$c^2$ mass, was called upsilon, $\Upsilon$, by it's
discoverers.  
There was evidence in the initial data of some excited states of the
ground-state resonance.  These were quickly confirmed at DESY in
Hamburg,~\cite{confirmY} where the DORIS storage ring energy was boosted
to be able to produce the new states.  Thus, the bottom quark came to
be an accepted member of the hierarchy of quarks.

As an aside, it might be noted that the $\tau$, the charged lepton of
the third generation, appeared just before the upsilon particle.  The
tau was not widely accepted, though it's discoverers were happy to see
the upsilon as a confirmation that the earlier picture of two
generations was incomplete.  The tau was, then, also soon widely
accepted.  Direct observation of the tau neutrino has, by the way, also
only just been announced this month.

A sixth quark was anticipated to be roughly 3 (or $\pi$) times the
mass of the bottom quark.  After all, there is such a pattern apparent
among the strange, charm, and bottom quarks. The TRISTAN accelerator in
Japan was even built with that goal in mind.  However, in spite of major
efforts, the discovery of the sixth, the ``top'' quark, did not occur
until twenty years later, in 1997.  The discovery required the dedicated
running of the highest energy colliding-beams accelerator in the world,
Fermilab's Tevatron Collider.  This was because the top quark was not
three times as heavy as the bottom quark, but about forty times as
heavy; weighing in at about 175 GeV/$c^2$.  The reason for this enormous
mass remains a mystery.

As the number of quarks grew, another feature of quarks appeared.  That
is, the eigenstates relevant to their production in strong and
electromagnetic interactions (called flavor eigenstates) are not the
same as the weak-interaction eigenstates of their decay (called mass
eigenstates).  The various eigenstates mix, as related by the 
Cabibbo-Kobayashi-Maskawa (CKM) matrix.~\cite{ckm}

\begin{equation}
Weak Eigenstates = {\bf V}  \times Flavor Eigenstates
\label{ckmmatrix}
\end{equation}
That is,  \\
$ \hspace*{4.0cm} d' \hspace{1.55cm}
V_{ud} \hspace{0.45cm} V_{us} \hspace{0.45cm} V_{ub} 
 \hspace{0.95cm} d$  \\
$ \hspace*{4.0cm} s' \hspace{0.5cm} = \hspace{0.5cm} V_{cd}
 \hspace{0.5cm} V_{cs} \hspace{0.5cm} V_{cb} \hspace{1cm} s$  \\
$ \hspace*{4.0cm} b' \hspace{1.6cm} V_{td} 
 \hspace{0.5cm} V_{ts} \hspace{0.5cm} V_{tb} \hspace{1.05cm} b$  \\
\\
Using the Wolfenstein parameterization,~\cite{wolfenstein}
to order $\lambda^3$, the CKM matrix is \\
\\
$\hspace*{3.2cm} 1 - \lambda^2/2  \hspace{2.5cm}   \lambda   
\hspace{2.0cm} A \lambda^3(\rho -{\it i}\eta) $\\
              
$\hspace*{0.3cm} {\bf V}  \hspace{0.1cm}  =   \hspace{1.9cm}  - \lambda
\hspace{2.4cm} 1 - \lambda^2/2  \hspace{1.9cm} A \lambda^2 $\\

$\hspace*{2.0cm} A \lambda^3(1- \rho- {\it i} \eta)  \hspace{1.5cm}  
-A \lambda^2  \hspace{2.4cm} 1 $ \\


The strong and electromagnetic interactions conserve the new quantum
numbers, called ``flavor:'' strangeness, charm, bottom, and top.  Thus,
production of these quarks in strong and electromagnetic interactions
always occurs in pairs; i.e., a quark and an antiquark of the same
flavor.  This is the origin of the ``strange'' behavior of the strange
particles -- and of the heavier quarks which followed.  On the other
hand, the weak interactions do not conserve flavor.  The violation of
flavor conservation occurs in a well defined way, however, with the
rates governed by the CKM matrix elements, $V_{ij}$, above.  

\subsection*{A Few Comments on Names}

The original names, up and down, came from an analogy with spinors,
with spin direction pointing up and down.  The ``strange'' quark name
was chosen as a reminder that it was supposed to explain the strange
behavior of particles containing these quarks; e.g., the $K$ mesons
which were always produced in pairs or in association with strange
baryons. Stranger still, even with strange quarks, some neutral $K$ 
meson decay behavior was not explained.  The branching ratios to certain
decay modes (e.g., $\mu^+\mu^-$) were much smaller than expected.
One proposed solution, the GIM mechanism~\cite{gim}, suggested a fourth
quark to fix things up.  This fourth quark was to have just the right
properties, be ``charmed'' in just the right way for the fix to work.
The name ``charm'' was actually suggested earlier by Bjorken and Glashow
in a  paper generalizing SU(3) quark symmetry to SU(4).~\cite{bjg}  I
think that the name stuck, in part, because of the charmed properties of
the quark.  In any event, the mass of the charm quark was predicted on
the basis of the needed properties for its cancellation of other
contributions to the neutral $K$ decays. 

The third generation returned to more prosaic names, ``top'' and
``bottom,'' like up and down,  but not before flirting with the names 
``truth'' and ``beauty.''  I have preferred these latter names, since I
used to describe my personal research as the ``search for truth and 
beauty'' -- but that was before they were both discovered!  This
takes us to near the end of the current story.  Let's review how the
heavy quarks were first observed, their properties, and the basic 
physics associated with each of them.

\section*{Experiment Techniques Leading to Heavy Quark Capabilities}

Heavy quarks have been studied in a large number of experiments at quite
a range of energies, from near threshold to much higher energies.
Nevertheless, it will be evident as each heavy quark is discussed that
certain experiment techniques have been critical to the success of the
physics program.  The same techniques appear and reappear.  They will
also be important for future physics efforts beyond heavy quarks.

The most important techniques for heavy quarks have been (1) efficient
event selection, (2) long data-taking runs with lots of beam,
(3) large data sets and lots of computing, and (4) the use
of solid-state detectors for precision charged-particle tracking.
Each of these will be discussed in turn, though it is the combination
of all of them together which has really made the difference.

\subsection*{Open, Efficient Event Selection}

For charm, the first extensive open charm particle studies were done
at $e^+e^-$ colliders running at the $\psi$(3S), also called the
$\psi$''.  Theses excited states of $c\overline{c}$ are produced
copiously relative to an underlying continuum of states, and decay
dominantly to $D$ $\overline{D}$ mesons.  Thus, experiments are
able to record all the hadronic events in their studies of the $D$
mesons.  Final event selections are made later, off-line.
 
The early attempts to study charm particles at fixed-target
experiments were less successful.  They attempted to use special
geometries for a single decay mode~\cite{fitch} and specialized triggers
for particular production mechanisms~\cite{e516}.  Later experiments
used looser event selection, and were more effective.  Amid the
generally-discouraging first efforts at fixed-target experiments, a list
of algorithms was examined for event selection.~\cite{appeltrigger}
The more successful experiments made their event selections based on
very open, efficient, inclusive triggers using total event transverse
energy and evidence for decays of particles with lifetimes in the
picosecond range.  The first of these was effective to the extent that
the new particles were heavy as a fraction of the center-of-mass energy,
the second to the extent that the new particles had long lifetimes or
decayed to particles with long lifetimes.

As was the case for charm, the first extensive studies of bottom
particles were most successful at $e^+e^-$ colliders.  For bottom
particles, the accelerators run at the energy where $\Upsilon$(4S) 
particles are produced.  Theses excited states of $b \overline{b}$
are produced copiously relative to an underlying continuum of
states, and decay dominantly to $B$ $\overline{B}$ mesons.  Thus,
experiments were again able to record all the hadronic events in their
studies of the $B$ mesons.  The openness of the on-line event selection
led to the ability of each experiment to attack a broad range of decays
and of physics topics.

Even for the top quark discovery, the on-line trigger is quite efficient
for top-quark events.  The properties of the top quark are so extreme,
that enormous numbers of less interesting events could be
discarded at the on-line trigger level.
  
\subsection*{Large Data Sets and Lots of Computing}

For each technique, technical progress has been critical to the
extensive progress made in heavy-quark physics.  In the case of data
acquisition and analysis, the progress has been adopted from the
commercial world where the cost per unit of data storage and computation
has dropped amazingly over the last two decades.  Table \ref{tableda}
lists the growth of charm samples achieved by increasing use of
parallelism and computer power per CPU in one series of experiments,
those at Fermilab's Tagged Photon Laboratory.  The basic experiment
apparatus (a forward, two-magnet spectrometer) did not change much
after the addition of silicon microstrip detectors in E691.  
Nevertheless, the number of reconstructed charm decays used in final
physics publications grew exponentially.  The physics signals shown in
the table improve by a factor of 2,000, not counting the improvement in
the signal-to-background ratio.  Such numbers may be taken as a rough
measure of the physics reach of the experiments.

\begin{figure}[b]
\vspace{2.5in} 
\epsfbox[0 0 30 50]{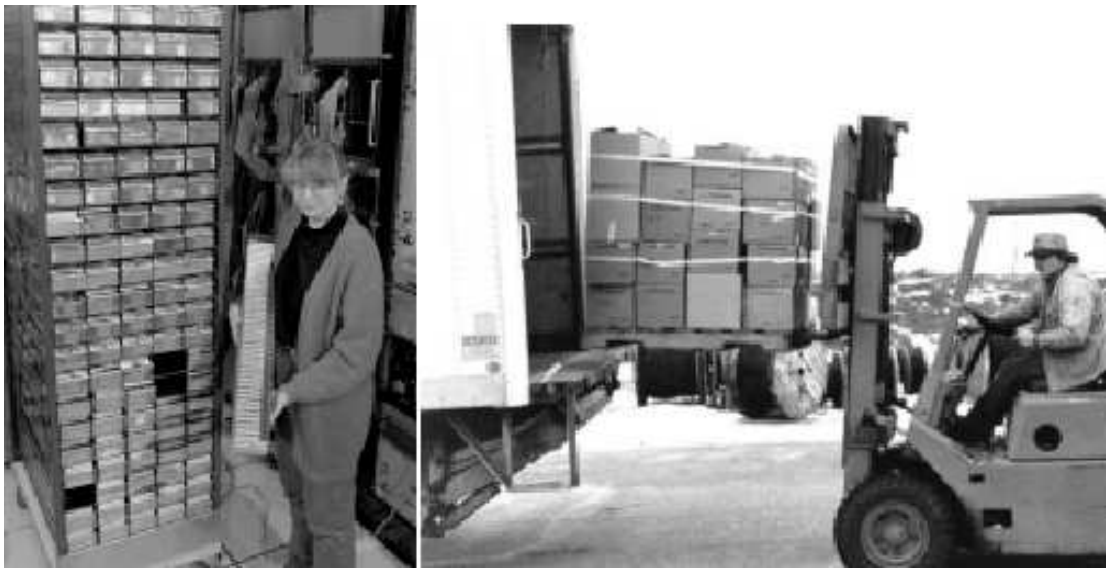}
\vspace{0.3cm}
\caption{A forklift arrives at E769 (right) after a weekend of data
taking using 9 track, 6250 bpi, open reel magnetic tapes.  E791
physicist Cat James (left) holds an arm-load of 8 mm video 
tapes next to a storage rack of such tapes.  An arm-load of tapes
was filled in parallel by 42 tape drives in less than three hours. }
\label{tpldata}
\end{figure}

The improved numbers of observed particles followed very nearly the
increases in data set size, which was made
affordable by the change from 9-track open-reel 6250 bpi magnetic tapes
of E769 and earlier, to the use of 8 mm video tape in E791.  A graphic
demonstration of this difference is given by comparing the two images in
Fig. \ref{tpldata}.  A fork lift and truck arrived at the E769
experiment each Monday morning to take the weekend's data tapes
to the computing center.  Compare that with nearly the same amount of 
data being held in the one arm-load of 8 mm tapes held by one of the
E791 physicists.  Perhaps even more impressive is the fact that the
arm-load of tapes were filled with data in just three hours, not a full
weekend.  A similar growth of efficiency and cost effectiveness was
needed in offline computing. It occurred via the use of ``farms'' of
cheap, parallel, networked CPUs.~\cite{nash}  Table \ref{expcdata} gives
the current status of the numbers of equivalent background free signals
in some representative charm experiments, calculated from quoted signal
sizes and their errors.

\begin{table}[t!]
\caption{Example of the growth of computing parallelism and power from
the series of charm experiments using the Fermilab Tagged Photon
Spectrometer.}
\centering   
\begin{tabular}{|c|c|c|c|c|c|c|c|}
\hline
Time  & Exp. & \# Data  & \# DAQ & \# Output & \# Rec'd. & Data Set
&
\# Reconst'ed \\  
Frame & \#   &  Streams &  CPUs  & Streams   & Events   &  Size    &  
Charm Decays  \\
      &      &          &        &           &($\times 10^6$)&(Gbytes) &
              \\
\hline
1980-2 &  E516 &  1 &  1 &  1 &     20 &     70 &     100 \\
1984-5 &  E691 &  2 &  1 &  1 &    100 &    400 &  10,000 \\
1987-8 &  E769 &  7 & 17 &  3 &    400 &   1500 &   4,000 \\
1990-2 &  E791 &  8 & 54 & 42 & 20,000 & 50,000 & 200,000 \\
\hline
\end{tabular}   
\label{tableda}
\end{table}

\begin{table}[b!]
\centering
\caption{ \it Numbers of Equivalent Pure Decays Observed by Analysis
(scaled to full data sets where needed).}
\begin{tabular}{|l|c|c|c|c|c|} 
\hline
Physics & Decay Mode  & ALEPH & E791 & CLEO II.V & FOCUS \\
 Topic  &             &       &      &           &       \\
\hline\hline
Mixing       &$D^o_{tag} \rightarrow K\pi
$&$1000$&$5,400$&$16,000$&
\\
             &$D^o_{tag} \rightarrow K\pi\pi\pi$&  &$3,300 $&       &
\\
\hline
Mixing       &$D^o_{tag} \rightarrow K\mu\nu $&    &$  750 $&       &
$7,400$\\
             &$D^o_{tag} \rightarrow Ke\nu   $&    &$  760 $&       &
\\
\hline
$\Delta\Gamma$&$D^o \rightarrow KK           $&    &$3,150 $&$ 1,700$&
        \\
             &$D^o \rightarrow K\pi          $&    &$29,500$&$30,000$&
$86,000$\\
             &$D^o \rightarrow K_s^o\phi     $&    &$      $&$ 4,100$&
\\
\hline
CP Vio.      &$D^+ \rightarrow KK\pi         $&    &$1,250 $&        &
$ 8,100$ \\
             &$D^+ \rightarrow \phi\pi       $&    &$  800 $&        &
\\
             &$D^+ \rightarrow K^*(890)K^+   $&    &$  420 $&        &
\\
             &$D^+ \rightarrow \pi\pi\pi     $&    &$  590 $&        &
\\
             &$D^+ \rightarrow K\pi\pi       $&    &$36,000$&        &
$120,000$\\
\hline
CP Vio.     &$D^o_{tag} \rightarrow K K      $&    &$  440 $& $2,100$&
$ 2,200$ \\
             &$D^o_{tag} \rightarrow \pi\pi    $&  &$  190 $&        &
\\
             &$D^o_{tag} \rightarrow K\pi\pi\pi$&  &$3,000 $&        &
\\
             &$D^o_{tag} \rightarrow K\pi      $&  &$10,500$&$13,500$&
$35,000$ \\
\hline
\end{tabular}
\label{expcdata}
\end{table}

\subsection*{Silicon Microstrip Detectors and Charge-Coupled-Devices}

\begin{figure}[b!]
\vspace{3.9in} 
\epsfbox[-100 0 30 50]{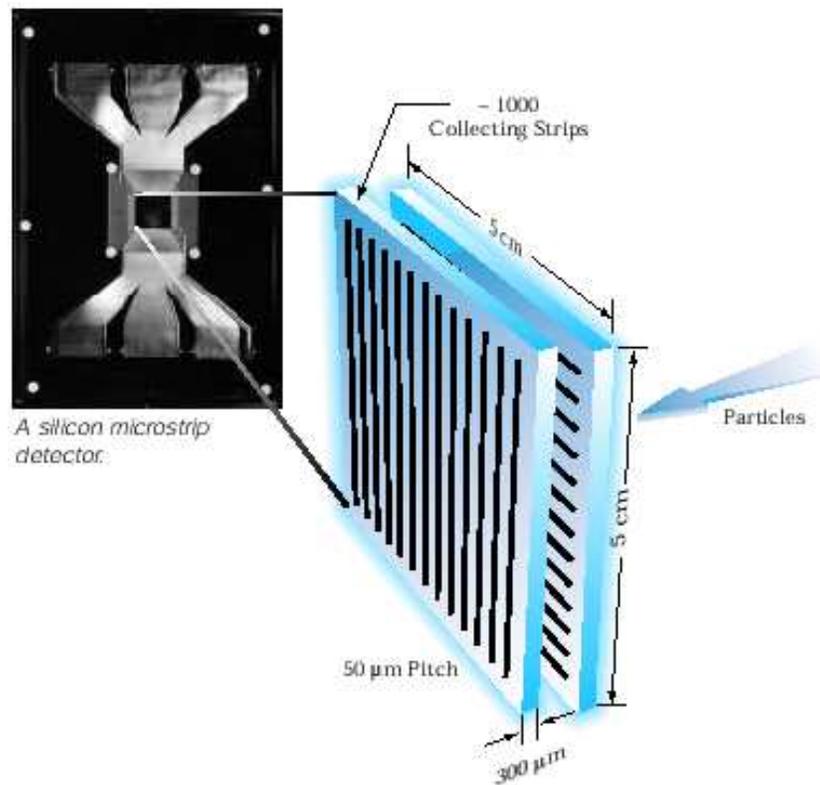}
\vspace{10pt}
\caption{A silicon microstrip detector and sketch showing how
orthogonal strips are used to map the trajectory of an incident
charged particle.  The solid-state detectors are fully depleted by
application of a reverse bias across their 300 micron thickness, with
signals of about 26,000 electrons observed.} 
\label{smd}
\end{figure}

\begin{figure}[b!]
\vspace{2.1in}
\epsfbox[93 297 268 337]{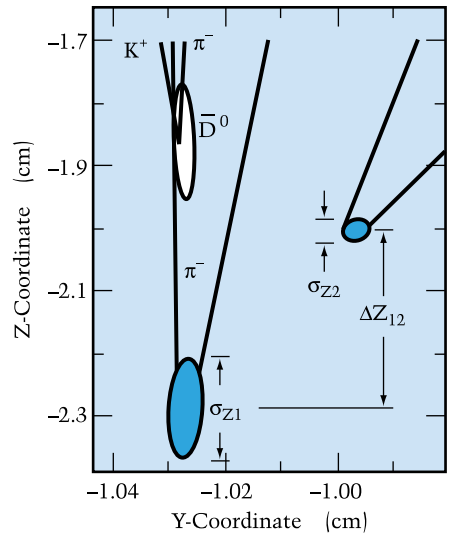}
\vspace{10pt} 
\caption{The reconstruction of vertices from trajectory information in
an example E691 photoproduction event.  The ovals represent the
uncertainty in the vertex fits.  Separation of the primary interaction
and the decay points of two charm particles are clearly evident.}
\label{vtx} 
\end{figure}

Silicon microstrip detectors~\cite{ACCMORsmd} (SMDs, shown in
Fig. \ref{smd}) and charge-coupled-devices (CCDs) provide very high
precision information about the trajectories of charged particles.  From
these trajectories, we can obtain the locations where the trajectories
overlap, e.g., the primary interaction point (primary vertex).  Such a
reconstruction from the charm photoproduction experiment E691 is 
shown in Fig. \ref{vtx}.  We see a primary vertex and two secondary
vertices where charm particles decay.  Given the decay of the two
long-lived particles near the primary interaction, the same tracking
devices find the decay location as well as the primary vertex.  How
long lived must be the particles for these observations to be made in the
laboratory?  The time scale is a picosecond.  A particle with such a
lifetime will travel 300 microns in the laboratory if its velocity is
such that $\beta\gamma$ is equal to 1.0.  This is only a little more
than typical longitudinal position resolution values.  So, experiments
do best when the particles are traveling with higher velocities in the
laboratory, and have larger values of $\beta\gamma$ -- reaching values
of a few tens even.  For fixed-target experiments with incident
charged-particle beams, one can also include the incident beam
track and thin-target locations in fits to find the best estimate
of the primary vertex location.

Use of the vertex information in an event provides a double benefit.
First, events with evidence of a secondary vertex near the interaction
point are highly enriched in heavy quark production.  Once a decay
vertex is located, the second benefit of precision tracking is
evident.  When searching for the right combination of observed
particles from a single decay, one need only examine the effective mass
of those particles coming from that decay vertex, not try all the
combinations of all particles in the event.  This reduces the
backgrounds due to random combinations of tracks by very large factors.

\section*{Observations and Physics of Charm Quarks}

\subsection*{Observations of Charm}

Although one long-lived event in cosmic rays\cite{niu} predated the
observation of the $J/\psi$, it was the observation of the $J/\psi$ in
both hadronic and $e^+e^-$ interactions that led to the wide acceptance
of charm.  The hadroproduction resulted in $e^+e^-$ pairs seen in the
experiment of Sam Ting and his group at Brookhaven.   At SLAC, the group
of Burton Richter saw a huge enhancement in the annihilation rate when
the center-of-mass energy of the colliding $e^+e^-$ beams was 3.1
GeV.  Given that the $J/\psi$ was made of a new quark-antiquark pair, a
host of other new particles containing such quarks was expected.  Among
the new particles were excited states of the quark-antiquark pair.  Some
of these are listed in Table \ref{charmonium}.  These so-called
charmonium states have no net charm, only ``hidden charm.''  We also
expected meson-combinations of charm quarks with lighter quarks and
baryon-combinations with three quarks, one or more of which carried
charm.  The particles with net charm are  known as ``open charm''
particles, and examples are listed in Table \ref{cparts}.

The early developments in charm physics were dominated entirely by
experiments at $e^+e^-$ colliders.  The fraction of events with charm
particles was large at the $\psi$''.  The backgrounds could be well
handled for many decay modes, even though it was necessary to examine
all combinations of tracks.  Once silicon microstrip detectors were
introduced into fixed-target experiments, combined with the other
features discussed above, the most precise measurements came to be
dominated by the Fermilab fixed-target program.  Now, we can see the
leadership position transferring to the asymmetric $e^+e^-$ collider
$B$ factories.  This is already evident in the preliminary results just
announced at the ICHEP2000 Rochester Meeting held the previous week in
Osaka, Japan.  Perhaps, eventually, the lead will pass to the forward
hadron-collider experiments at Fermilab and CERN -- BTeV and LHC-b,
respectively.  Given the plans evident in the community today, future
charm physics will come as a byproduct of the more intensive efforts
aimed at understanding the $b$-quark system.

\subsection*{Techniques in Charm Experiments}

The techniques discussed earlier used many charm experiment examples.
They will not be repeated here.  It is worth noting some additional
features, nevertheless.  Among these are the so-called $D^*$-trick, the
use of the beam energy in making mass plots of charm candidates from
$\psi''$ data, the use of additional kinematic and topological criteria,
the use of high $p_t$ leptons in triggers at fixed-target experiments,
and charged particle identification as examples.

One special kinematic feature in charm is the very small kinetic energy
available in the decay of the $D^*$, the first excited state of the 
$D$ meson.  This has been a useful feature, not only to find these $D^*$
mesons in complicated environments like high-energy colliders, but also
in selecting events where the nature at birth of the decay $D^o$ is
known.  Thus, events with a $\pi^+$ which comes from a $D^{*+}$ decay
means that the accompanying $D^o$ is not a $\overline{D}^o$.  One can
examine such events for evidence of mixing, that is if the $D^o$ has
become a $\overline{D}^o$ before it decays. 

At $e^+e^-$ colliders running at the $\psi''$, backgrounds are reduced
and kinematic parameter resolution improves using the beam energy in
making mass spectra distributions. The mass calculated this way is
called the ``beam constrained mass.''  The requirements used and the
calculations are:

\begin{equation}
\Delta E = E_{candidate} - E_{beam}
\end{equation}
and     
\begin{equation}
M_{B \ candidate} = \sqrt{E_{beam}^2 - p_{candidate}^2}
\end{equation}

In addition to looking for separations of decay vertices from the
location of the primary interaction, one can demand for fully charged
decay modes that the vector sum of the decay products point back
to the primary vertex.  This helps reduce backgrounds due to false
combinations of reconstructed tracks.  The technique even works, though
less well, with a missing particle -- especially if that particle is
light as in the case of a missing neutrino.  Of course, one must allow
for some mismatch.  One can also get to a two-fold-ambiguous momentum of
a missing particle if one assumes the mass of that particle as well as
the parent particle.  The farther from the production point the decay
is, the better this reconstruction performs.

\subsection*{Spectrum of Charm Mesons and Baryons}

The first particles containing charm quarks to be observed were the
so-called ``onium'' states of a charm quark and charm antiquark.  These
states have no net ``charm,'' and can decay electromagnetically and, for 
the heavier such states, strongly.  Thus, widths of the lower mass
states are very narrow, and signals appear clearly above background.
On the other hand, such states have unmeasurable decay lengths in the
laboratory, making some observations difficult, if not impossible.
Some of the observed charmonium states are listed in Table
\ref{charmonium}.  Study of the charmonium states is still an active
area.  The $\eta_c'$, for example, has not had a confirmed observation
yet, in spite of several attempts.  In addition, there are many more
excited states of these onium particles.

\begin{table}[t!]
\caption{Examples of $c\overline{c}$ hidden-charm particles, their
masses, widths, and typical decays.}
\centering
\begin{tabular}{|c|c|c|c|c|}
\hline
$J^{PC}$ &    Symbol      &       Mass         &    Width           &
Decay Mode                   \\
Assignment&               &    (MeV/$c^2$)     &  (MeV/$c^2$)       &
Examples                     \\
\hline\hline
$0^{-+}$ &$\eta_c$(1S)    &$  2980 \pm 2      $&$ 13.2^{3.8}_{3.2} $& 
  $ \eta\pi\pi$              \\  
\hline
$0^{-+}$ &$\eta_c$(2S)    &                     &                    &    
  not observed               \\
         & $\eta_c'$      &                    &                    &
                             \\ 
\hline
$1^{--}$ &$ J/\psi$(1S)   &$ 3096.87 \pm 0.04 $&$ 0.087 \pm 0.005 $&
  $ 2(\pi^+\pi^-)\pi^o$      \\
\hline
$1^{--}$ &$  \psi$(2S)    &$ 3685.96 \pm 0.09 $&$ 0.277 \pm 0.031 $&
  $J/\psi$(1S)$\pi^+\pi^-$   \\
         &$  \psi'       $&                    &                   &
  $J/\psi$(1S)$\pi^o\pi^o$   \\   
\hline
$1^{--}$ &$\psi$(3S)    &$ 3769.9  \pm 2.5  $&$ 83.9 \pm 2.4    $& 
  $ D \overline{D} $         \\
         &$\psi'''       $&                   &                    &
                             \\
\hline
$0^{++}$ &$\chi_{c0}$(1P) &$  3415.0 \pm 0.8 $&$ 14.9^{2.6}_{2.3} $&   
  $  2(\pi^+\pi^-)   $        \\
\hline
$1^{++}$ &$\chi_{c1}$(1P) &$  3510.5 \pm 0.1 $&$  0.88 \pm 0.14   $&    
  $ \gamma J/\psi$(1S)        \\
\hline
$2^{++}$ &$\chi_{c2}$(1P) &$  3556.2 \pm 0.1 $&$  2.0 \pm 0.2     $& 
  $ \gamma J/\psi$(1S)        \\
\end{tabular}
\label{charmonium}
\end{table}

It is the ground states with open charm that have longer lifetimes, and
have been amenable to selection via their observable laboratory flight
paths. The ground-state, open, single-charm mesons and baryons are
listed in Table \ref{cparts}.  No baryons with two charm particles among
their three quarks have yet to be observed, though they should certainly
appear eventually.  Each of the ground-state particles can have a set of
excited states, due either to radial or angular-momentum excitations.
Many of these have already been observed, and there is a long story to
be told here.  Let me just note that these states are well described by
focusing on the heavy charm quark as defining the coordinates, with an
antiquark or diquark system orbiting around it.  Spectrum mass-level
separations agree with predictions from lattice gauge QCD calculations,
and one can now even obtain values for the strong coupling constant from
the level splittings in these states.

\begin{table}[t!]
\caption{Examples of open-charm particles and their decays.}
\centering
\begin{tabular}{|c|c|c|c|c|}
\hline
Quark                  &  Symbol   &    Mass           &
 Lifetime  &       Decay Mode                       \\
Combination            &           & (MeV/$c^2$)       &
  (fs)     &       Examples                       \\
\hline
${\it c}\overline{\it d}$& $ D^o $   &$ 1864.5 \pm 0.5 $ &  
$ 413 \pm 2.8$  &   $K^-\pi^+,K^-\pi^-\pi^+\pi^+$    \\
${\it c}\overline{\it u}$& $ D^+ $   &$ 1869.3 \pm 0.5 $ &
$1051 \pm 1.3$  &   $K^-\pi^+\pi^+$                  \\
${\it c}\overline{\it s}$& $ D_s^+      $&$ 1968.6 \pm 0.6 $ &
$496^{+10}_{-9}$&   $K^-K^+\pi^+ $                   \\
${\it c}{\it u}{\it d}$&$\Lambda_c    $&$ 2284.9 \pm 0.6 $ &
  $ 206 \pm 12 $&  $ p K^- \pi^+ $                   \\
${\it c}{\it u}{\it u}$&$\Sigma_c^{++}$&$ 2452.8 \pm 0.6 $ &
                &  $ \Lambda_c^+ \pi^+ $             \\   
${\it c}{\it u}{\it d}$&$\Sigma_c^+   $&$ 2453.6 \pm 0.9 $ &
                &  $ \Lambda_c^+ \pi^o $             \\   
${\it c}{\it d}{\it d}$&$\Sigma_c^o   $&$ 2452.2 \pm 0.6 $ &
                &  $ \Lambda_c^+ \pi^- $             \\
${\it c}{\it s}{\it u}$&$\Xi_c^+      $&$ 2466.3 \pm 1.4 $ &
  $ 330^{+ 60}_{- 40} $ & $ \Lambda K^- \pi^+ \pi^+ $\\
${\it c}{\it s}{\it d}$&$\Xi_c^o      $&$ 2471.8 \pm 1.4 $ &
  $  98^{+ 23}_{- 15} $ &   $\Xi^- \pi^+,\Omega^-K^+$\\
${\it c}{\it s}{\it s}$&$\Omega_c     $&$ 2704 \pm 4     $ &
  $  64 \pm 20        $ & $ \Sigma^+ K^- K^- \pi^+  $\\
\end{tabular}
\label{cparts}
\end{table}

\subsection*{Physics Issues for Charm Quarks}

The physics issues for charm quarks range from searches for clues to new
physics to contributions to the understanding and parameters of the
Standard Model.  The searches for physics beyond the Standard Model
include those which would appear as CP violation, oscillations between
neutral states, and searches for forbidden and overly copious rare
decays.  Among the Standard Model parameters to be measured accurately
are the CKM matrix elements: $|V_{cs}|$ and $|V_{cd}|$,
determination of the strong coupling constant $\alpha_s$ as a test
of the flavor independence of QCD, branching ratios and lifetimes to
high precision, and a myriad of resonance and other non-perturbative
parameters.

\subsection*{Charm, a Unique Window to New Physics}
        
Two features of charm make it unique in the search for new physics:
(1) the absence of Standard Model background and (2) the fact that
coupling to charm is the only way to see new physics in the up-quark
sector.  The Standard Model sources of mixing and CP violation for
strange and bottom quarks predict significant, even large effects.  Yet,
for charm, such effects are predicted (so far) to be unmeasurable. Thus, 
experimental signatures in charm have no SM background, no relevant
hadronic uncertainty in background estimates.  Any sign of mixing or
CP violation in the charm sector is immediate evidence of new
physics.~\cite{burdman,liu}

As for new physics coupling to the up-quark sector, both the up-quark
and the top-quark are prevented from having observable effects in
virtually all the standard ways of looking.  There is a lack of
decay channels for the up quark itself. So, interferences among decay
channels are severely constrained.  The top quark doesn't live long
enough to mix or have the final state interactions needed for CP
violation.

Mixing in the Standard Model is a good example of why effects are so
small in the charm sector.  The Standard Model process is thought to be
dominated by contributions of the so-called ``box diagram'' shown
below.  Down-type quarks appear in the loop for this mechanism.  The
amplitude has the form \\
\begin{equation}
    A_{mix}($charm box diagram$) \sim (m_s^2 - m_d^2)/ m_W^2 \times
(m_s^2 - m_d^2)/ m_c^2
\end{equation}
where the first term comes from the sum of the two leading terms (GIM
suppression), and the second is an off-shell factor.  In the limit of
SU(3) symmetry, the down and strange quarks have equal masses and
the amplitude is zero.  Comparing\\

\input feynman
\begin{picture}(18000,24000)
\Large
\THICKLINES
\startphantom
\drawline\photon[\E\REG](0,18000)[8]
\global\Xeight=\plengthx
\stopphantom
\drawline\fermion[\E\REG](8000,18000)[\Xeight]
\global\Xone=\fermionfrontx
\global\advance\Xone by -2000
\put(\Xone,\pmidy){Q}
\drawline\photon[\S\REG](\fermionbackx,\fermionbacky)[8]
\global\Xone=\pmidx
\global\Yone=\pmidy
\global\advance\Xone by -1500
\put(\Xone,\Yone){\makebox(0,0){W}}
\drawline\fermion[\W\REG](\photonbackx,\photonbacky)[\Xeight]
\global\Xone=\fermionbackx
\global\advance\Xone by -2000
\put(\Xone,\pmidy){$\overline{\mathrm{q}}$}
\global\Xtwo=\pbackx
\global\Ytwo=\pbacky
\global\advance\Ytwo by -5000
\put(\Xtwo,\Ytwo){\makebox(0,0)[l]{Q = \hspace{0.05cm} c \hspace{1.25cm}
--
\hspace{1.0cm} $Q_{loop}$ = b,s,d}} 
\global\advance\Ytwo by -2000
\put(\Xtwo,\Ytwo){\makebox(0,0)[l]{Q = s,b \hspace{1.0cm} -- 
\hspace{1.0cm} $Q_{loop}$ = t,c,u}}
\message{Xtwo=\the\Xtwo}  
\message{Ytwo=\the\Ytwo}
\drawline\fermion[\E\REG](\photonbackx,\photonbacky)[\Xeight]
\global\Xone=\pmidx 
\global\Yone=\pmidy
\global\advance\Yone by -1500
\put(\Xone,\Yone){\makebox(0,0){$Q_{loop}$}}
\drawline\fermion[\E\REG](\photonfrontx,\photonfronty)[\Xeight]
\global\Xone=\pmidx
\global\Yone=\pmidy
\global\advance\Yone by 1500
\put(\Xone,\Yone){\makebox(0,0){$Q_{loop}$}}
\drawline\photon[\S\REG](\fermionbackx,\fermionbacky)[8]
\global\Xone=\pmidx
\global\Yone=\pmidy
\global\advance\Xone by 1500
\put(\Xone,\Yone){\makebox(0,0){W}}
\drawline\fermion[\E\REG](\photonbackx,\photonbacky)[\Xeight]
\global\Xone=\fermionbackx
\global\advance\Xone by 2000
\put(\Xone,\pmidy){$\overline{\mathrm{Q}}$}
\drawline\fermion[\E\REG](\photonfrontx,\photonfronty)[\Xeight]
\put(\Xone,\pmidy){q}
\end{picture}
%
\\[0.3cm]
this to the same sort of diagram for
mixing of neutral kaons with up-type quarks in the loop shows the
orders-of-magnitude difference due to the coupling constants (CKM
parameters) and the masses of the quarks involved.
\begin{equation}
    A_{mix}($kaon box diagram$) \sim (m_c^2 - m_u^2)/ m_W^2 \times
(m_s^2 - m_d^2)/ m_s^2.  
\end{equation}
Note that for neutral kaon decays to dileptons to be as small as they
are, the cancellation of the charm and up quark contributions is
required in such loop diagrams.  This is what allowed Glashow,
Iliopoulos, and Maiani to predict  the mass of the charm quark before
there was any direct evidence for it.
Contributions from these box diagrams, and even those from penguin
and long distance effects are of the order of $10^{-7}$ to $10^{-10}$.
A rather complete compilation of model-dependent calculations for
both Standard-Model and new-physics models is being maintained by Harry
Nelson.~\cite{nelson}

We often refer to the gap between Standard Model backgrounds and the
current level of experiment limits as an open window.  That is, there is
an opening for a major discovery.  Table \ref{charmnewphys} lists a
number of measurements, the current levels of sensitivity, and the
theoretical estimates of the contributions of Standard Model sources to
possible signals.  There are orders of magnitude available in these open
windows.  Furthermore, there are many extensions of the Standard Model 
which might appear in the window.  These include fourth-generation
quarks, leptoquarks, and various other heavy particles (e.g., Higgs and
SUSY particles) which can appear in loops in virtual processes.

\begin{table}[t!]
\caption{Exmples of the Open Window Sensitivity to Physics Beyond the
Standard Model.}
\begin{center}
\begin{tabular}{|l|c|c|c|}
\hline
&  & SM  & Typical\\
\raisebox{1.5ex}[0pt]{Topic} & \raisebox{1.5ex}[0pt]
{ 90\% CL Limit}
& prediction  & Models Tested \\
\hline\hline
{\em CP} Violation
& & & \\ \hline
~$D^0\to K^- \pi^+$              & -0.009$<$$a$$<$0.027~\cite{Bartelt}
& $\approx0$ (CFD)               &  SUSY,      \\
~$D^0\to K^- \pi^+\pi^-\pi^+$    &
& $\approx0$ (CFD)               &  LR Symm.,  \\
~$D^0\to K^+ \pi^-$              & -0.43$<$$a$$<$0.34~\cite{cpcleo00}
& $\approx0$ (DCSD)              &  Extra Higgs\\
~$D^+\to K^+ \pi^+ \pi^-$        &
& $\approx0$ (DCSD)              & \\
~$D^0\to K^- K^+$                & -0.026$<$$a$$<$0.028\cite{E831cp}
&                                &                      \\
                                 & -0.093$<$$a$$<$0.073\cite{E791cpneut}
&                                &                      \\
                                 & -0.022$<$$a$$<$0.18\cite{Bartelt}
&                                &                      \\
~$D^0\to \pi^+\pi^-$             & -0.002$<$$a$$<$0.094\cite{E831cp}
&                                &                      \\
                                 & -0.186$<$$a$$<$0.088\cite{E791cpneut}
&                                &                      \\
~$D^+\to K^- K^+\pi^+$           & -0.006$<$$a$$<$0.018\cite{E831cp}
&                                &                      \\
                                 & -0.026$<$$a$$<$0.028\cite{E791cpchg}
&                                &                      \\
~$D^+\to \overline K^{*0} K^+$   & -0.092$<$$a$$<$0.072\cite{E791cpchg}
& $(2.8\!\pm\!0.8)\times 10^{-3}$\cite{Pugliese}     & \\
~$D^+\to \phi\pi^+$              & -0.075$<$$a$$<$0.21\cite{Frabetti}
&                                &                      \\
~$D^+\to \pi^+\pi^+\pi^-$        & -0.086$<$$a$$<$0.052\cite{E791cpchg}
&                                &                      \\
~$D^+\to \eta\pi^+$              &   
& $(-1.5\!\pm\!0.4)\times10^{-3}$~\cite{Pugliese}    &  \\
~$D^+\to K_S\pi^+$               &   
& few $\times 10^{-4}$ \cite{Xing} & \\
\hline
FCNC
& & & \\ \hline
~$D^0\to\mu^+\mu^-$              & $4 \times 10^{-6}$
~\cite{E771mumu,WA92mumu}
& $<3\times10^{-15}$~\cite{Hewett}                     & $4^{th}$
Gen.,\\
~$D^0\to \pi^0\mu^+\mu^-$        & $1.7\times10^{-4}$~\cite{E653}
&                                & Tree-level \\
~$D^0\to \overline K^0 e^+e^-$   & $1.1\times10^{-4}$~\cite{E687fcnc}
& $<2\times10^{-15}$~\cite{Hewett}                     & FCNC \\
~$D^0\to\overline K^0\mu^+\mu^-$ & $2.6\times10^{-4}$~\cite{E653}
& $<2\times10^{-15}$~\cite{Hewett}                     & \\
~$D^+\to \pi^+e^+e^-$            & $6.6\times10^{-5}$~\cite{E791ll}
& $<10^{-8}$~\cite{Hewett}       & \\
~$D^+\to \pi^+\mu^+\mu^-$        & $1.8\times10^{-5}$~\cite{E791ll}
& $<10^{-8}$~\cite{Hewett}       & \\
~$D^+\to K^+ e^+e^-$             & $2.0\times10^{-4}$~\cite{E687ll}
& $<10^{-15}$~\cite{Hewett}      & \\
~$D^+\to K^+ \mu^+\mu^-$         & $9.7\times10^{-5}$~\cite{E687ll}
& $<10^{-15}$~\cite{Hewett}      & \\
~$D\to X_u+\gamma$               &
& $\sim10^{-5}$~\cite{Hewett}    & \\
~$D^0\to \rho^0\gamma$           & $2.4\times10^{-4}$~\cite{fcnc-cleo}
&$(1-5)\times10^{-6}$~\cite{Hewett}                   & \\
~$D^0\to \phi\gamma$             & $1.9\times10^{-4}$~\cite{fcnc-cleo}
&$(0.1-3.4)\times10^{-5}$~\cite{Hewett}               & \\
\hline
LF or LN Violation\
& & & \\ \hline
~$D^0\to\mu^\pm e^\mp$ & $8.1\times 10^{-6}$~\cite{E791mu-e}    & 0
& LQ \\
~$D^+\to\pi^+\mu^\pm e^\mp$ & $3.4\times 10^{-5}$~\cite{E791mu-e} & 0
& \\
~$D^+\to K^+ \mu^\pm e^\mp$ & $6.8\times 10^{-5}$~\cite{E791mu-e} & 0
& \\
~$D^+\to \pi^- \mu^+\mu^+$ & $1.7\times 10^{-5} $~\cite{E791mu-e} & 0
& \\
~$D^+\to K^- \mu^+\mu^+$ & $1.2\times 10^{-4}$~\cite{E791mu-e}    & 0
& \\
~$D^+\to \rho^- \mu^+\mu^+$ & $5.6\times 10^{-4}$~\cite{E653}   & 0
& \\
\hline
Mixing
& & & \\ 
\hline
~${}^{^{(}}
{{\overline D}{}^{^{)}}}^0 \to K^\mp\pi^\pm$ &  
$\Delta M_D<2.8\!\times\!10^{-5}$\,eV ~\cite{cpcleo00}    & 
$10^{-7}$\,eV~\cite{burdman,Burdman}        & LQ, SUSY, \\
~${}^{^{(}}
{{\overline D}{}^{^{)}}}^0 \to K\ell\nu$ & $r<0.005$~\cite{E791slmx} & &
$4^{th}$ Gen.,  \\
  &  &  & Higgs \\
\hline
\end{tabular}  
\end{center}
\label{charmnewphys}
\end{table}

\subsection*{Charm Production - Cross Sections}

The charm cross section is relatively small for fixed-target
photoproduction, where charm is produced in only about a half percent of
hadron-producing interactions.  The charm cross section is even smaller
in fixed-target hadroproduction, where charm is produced only about once
per thousand interactions.  Once one gets to Tevatron Collider energies,
the fractional cross sections rise by an order of magnitude.  

The theoretical uncertainties associated with charm hadroproduction
predictions are about an order of magnitude at fixed-target energies,
even for next-to-leading-order calculations.  This is due to the
sensitivity to the charm quark mass and the scale dependence of these
calculations.  In principle, calculations for production at very high
transverse momentum might mitigate the scale dependence, but this has
yet to be observed.  Calculations for photoproduction have somewhat less
uncertainty.

The charmonium production is a small fraction of total charm 
hadroproduction.  The comparison with calculations for the $J/\psi$ and
$\psi$' are complicated by the existence of sources like bottom decay
and the decay of the higher excited charmonium states.   Nevertheless,
the levels of these sources have been measured.  Once these are
subtracted, a mystery remains.  At both fixed-target and collider
energies, the direct production  $J/\psi$ and $\psi$' mesons are
measured to be factors of 7 and 25 larger, respectively, than predicted
by the simple color-singlet model.  When this is ``explained'' by
including color-octet contributions, the hadron-collider matrix elements
are not consistent with the levels observed at the HERA $ep$ collider.

\subsection*{Charm Production - Hadronization Effects}

The process of produced charm quarks turning into the charm mesons and
baryons observed in the laboratory is called hadronization.  The process
is non-perturbative by its very nature.  Nevertheless, some patterns are
emerging.  For one thing, the longitudinal momentum of observed charm
mesons in hadronic interactons looks very much like the predicted
distribution calculated for quarks.  This appears to come about because
of a cancellation of ``color drag'' and fragmentation.  The first of
these is an acceleration of the produced quarks in the direction of the
incident particles.  This is said to be due to the pull of the color
strings which attach the charm quark to the forward-going remnants of
the incident particle.  The fragmentation is the deceleration of the
charm quarks as they pick up sea-quarks 
\\
\\
\\
to make observable particles.  
There is some evidence for the attachment of color strings to the charm
and valence quarks of projectiles.  This is called the ``leading
particle'' effect.  In this effect, we see an asymmetry in the forward
direction in the numbers of particles with valence quarks in common with
the projectile compared to those without.  For example, in incident 
negative-pion-beam experiments, there is a preponderance of $D^-$'s over
$D^+$'s.  For kaon beams, the preponderance is of $D_s$ mesons with the
same strange quark as the valence strange quark in the incident kaons.

The details of the production process are also probed by
measurements of the correlations of charm and anti-charm particles.  One
mystery here is why the charm and anti-charm particles do not appear
more nearly back-to-back in the plane transverse to the incident
particle.  Some smearing can come from the intrinsic transverse
momentum, $k_t$, of the partons in the incident particles.  However,
the amount of $k_t$ required to explain observations in hadroproduction
can be as much as 2 to 3 GeV/$c$.  This is rather large, even for
a nucleon whose total rest mass is on the order of 1 GeV/$c^2$.

\subsection*{Charm-Particle Lifetimes}

The lifetimes of charm particles would all be the same if they all were
the result of a single process, say the color-aligned spectator diagram
where the charm quark decays to the Cabibbo-favored $W^+$ and a strange
quark, with the $W^+$ becoming some final state particle via
its virtual decay to $u \overline{d}$. However, there are other
possibilities: an annihilation diagram, a $W$-exchange diagram, and 
a second (color suppressed) spectator diagram where the quarks from the
$W$ decay do not stay together.

Various of the mesons and baryons have differing contributions from each
of these diagrams.  From the lifetimes listed in Table \ref{cparts}, it
is evident that more than a single process must be relevant.  Lifetimes
of the ground-state charm particles vary by an order of magnitude!  A 
consistent picture of charm decay requires inclusion of all these
processes, including coherent interference of diagrams when the
final-state particles from two diagrams are the same.~\cite{cheung}
There is also evidence of enhancement of the hadronic modes by diagrams
with gluons.~\cite{bigi}  This gluon participation may account for the
fact that the color-suppression in charm decay is not as pronounced as
in bottom decays.
 
\subsection*{Charm Decay - Resonance Dominance and Light Resonance
Parameters}

\begin{figure}[b!] 
\vspace{4.7in}
\epsfbox[0 0 30 50]{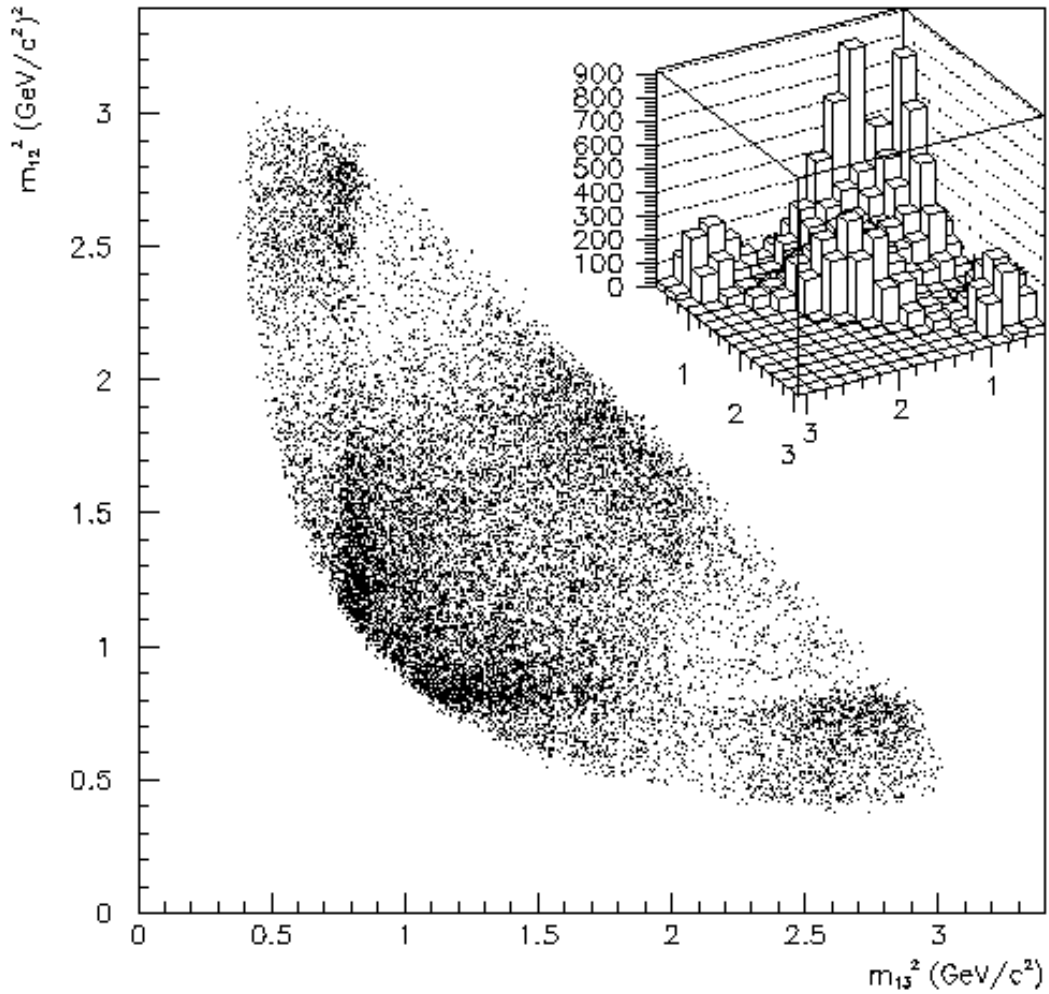}
\caption{The Dalitz distribution for the decay $D^+ \rightarrow 
K^- \pi^+ \pi^+$.  Since the two pions are indistinguishable, the data
has been symmetrized in making the figure.}
\label{Kpipi}
\end{figure}

Detailed study of the decays of charm mesons and baryons shows a 
dominance by quasi-two-body modes, that is where the charm particle
decays to a low-mass hadron resonance and another particle.  The study
of such decays is often pursued by examining the distribution of decays
in terms of their Dalitz plot, the two-dimensional distribution for
three particles in the final state which is uniform for non-resonant
s-wave decays, and has various structures evident for other modes.  An
example is the Dalitz plot for $K \pi \pi$ decay of the $D^+$ meson
shown in Fig. \ref{Kpipi}.  Each dot represents a measured decay; it's
location in the plot is a function of the location in the available
decay phase-space. 

The plot is a beautiful demonstration of basic physics, for example
that observations are driven by the square of the more fundamental
amplitudes.  We see that in the distinct interference pattern between
the $K^*(890)$ resonance, observed as a band at the $K^*$-mass squared,
and the more slowly varying other decay contributions.  We also see the
conservation of angular momentum in this decay of a pseudoscaler
particle into a vector particle, the $K^*$, and a pseudoscalar, the
recoiling pion in the decay.  This combination of spins, leads to a $(1
+ cos(\theta))$ decay angular distribution, which shows up in the Dalitz
plot as a concentration of $K^*$ events near the kinematic boundary.
Thus, we see that the distributions in the Dalitz plot are 
characteristic of the resonances involved and their masses, widths, and
spins.  

As is evident from this one Dalitz plot example, the distribution is
sensitive to which resonances contribute to the decay, the mass and
width of those resonances, and to the relative phase of the
contributions of the decay.  Much of the information on low-mass
resonances has come from scattering experiments.  There, the partial
waves which contribute can be quite complex.  In the decay of charm
particles, the initial state is somewhat better defined, less complex.
Thus, charm decay can be used as an independent way of understanding
light-resonance physics.  

One of the most compelling examples of this is in the three charged
pion decays of the $D^+$ and $D_s^+$ mesons.  In these decays, one
observes the contributions from such final states as $\rho^o \pi^+,
f_o(980) \pi^+, f_2(1270)\pi^+, f_o(1370)\pi^+, \rho^o(1450)\pi^+$, and
$\sigma \pi^+$.  The decays are dominated by the scalar plus $\pi^+$
modes, and the initial quark content of the D's and that of the
resonances can be understood to be consistent.  This latter point is
useful in determining the decay mechanisms at work (e.g., what
fraction of the decays come from the annihilation diagram) and the
nature of some of the less-well understood resonances such as the
$f_o(980)$, the $f_o(1270)$, and the $\sigma$.  The parameters of these
resonances have been determined from the fits to the relevant Dalitz
distributions.  Although it is not so clear from Fig. \ref{Kpipi}, a
very large part of the slowly varying density of decays comes from the
scalar $\kappa$ resonance. It's mass and width are being determined in a
new result from E791.\cite{E7913pi}

Study of the Dalitz plots of charm particles can also help in untangling
the contributions of the basic quark-level decay diagrams.  Various
resonances are characteristic of different decay mechanisms for a given
charm parent.  So far, the contributions of non-spectator diagrams for  
mesons appears to be somewhat limited.  However, baryons have a richer
set of possibilities since helicity is no longer a supressing
factor.  In fact, the variety of baryon lifetimes can be explained by  
incorporating the variety of possible quark-level diagrams.

\subsection*{Summary of Charm Physics}

We have gone very quickly over the enormous range of physics topics
where charm quark experiments can contribute.  These included
(1) new physics searches, (2) Standard Model electroweak parameters,
and (3) QCD understanding and parameters in production, spectroscopy,
and decay mechanisms.

\section*{Observations and Physics of Bottom Quarks}

As in the case of charm quarks and their physics, I will briefly review
the first observations of bottom, the types of experiments which have
studied bottom particles, the particular techniques used, and finally
the range of physics topics of special interest for bottom.

\subsection*{Observations of Bottom Quarks}

The first observation of bottom quarks was in a 1977 fixed-target
hadroproduction experiment at Fermilab.  There, the group headed by Leon
Lederman observed $\mu^+\mu^-$ pairs in a complicated peak near a mass
of 9.5 GeV/$c^2$ above a continuum.~\cite{discoveryY}  The
experimenters,
conscious of the resonances in the J/$\psi$ system, recognized that 
they might be seeing such a system for a new, third generation quark.
The Fermilab results led to an energy upgrade of the DORIS $e^+e^-$
collider at DESY in Hamburg, Germany.  There, several experiments were
able to offer rather quick confirmation of the
discovery.~\cite{confirmY} 
Because of the much better mass resolution based on beam energies at
DORIS, the experimenters were able to discern three resonances, just
where the Lederman group had said they were.

\subsection*{Spectrum of Bottom Mesons and Baryons}

Similarly to the case of charm, but with an even greater potential
variety of particles, the bottom system observations are all consistent
with the pattern expected for the usual meson and baryon quark
combinations.  Table \ref{bparts} gives examples of the better known
particles.  Many others are anticipated, but have yet to be observed.  
There are also many states of hidden ``beauty,'' not shown, and of
radial and angular-momentum excitations of all the ground states.  Even
for the observed particles, however, the full demonstration of their
expected spin and parity has yet to be accomplished.

\begin{table}[t!]
\caption{Examples of open-bottom particles and their decays.}
\centering
\begin{tabular}{|c|c|c|c|c|}
\hline
Quark           &   Symbol     &    Mass         &  Lifetime      &
Decay Mode      \\
Combination     &              &  (MeV/$c^2$)    &   (fs)         &
Examples        \\
\hline
 $\overline{b} d    $&$B_d$ or $B^o$&$  5279 \pm 2   $&$ 1548 \pm 32 $ &
$D^-\pi^+ \pi^+ \pi^-$\\
 $\overline{b} u    $& $ B^+   $    &$  5279 \pm 2   $&$ 1653 \pm 28 $ &
$      D^o \rho^+$    \\
 $\overline{b} s    $& $ B_s   $    &$  5369 \pm 2   $&$ 1493 \pm 62 $ &
$     D_s^-   X$      \\
 $\overline{b} c    $& $ B_c   $    &$ 6400 \pm 430$&$460^{+180}_{-160}$&
$    J/\psi \pi^+$    \\
   $  b u d         $&$\Lambda_b$   &$  5624 \pm 9   $&$ 1229 \pm 80 $ &
$   J/\psi\Lambda$    \\
\end{tabular}
\label{bparts}
\end{table}

\subsection*{Special Techniques Used for Bottom}

Particles containing bottom quarks have been observed in a great variety
of experiments, from threshold to very high energies -- more so than any
of the other heavy quarks.  At symmetric $e^+e^-$ colliders such as
CLEO at Cornell, the machines are operated at the $\Upsilon(4S)$ where
the resonance is just above $B \overline{B}$ threshold.  Thus, since the
center of mass is nearly at rest in the laboratory, the $B$'s are
also nearly at rest in the laboratory.  In order to produce $B$'s with
significant laboratory lifetimes, Pier Oddone and collaborators proposed
building an asymmetric $e^+e^-$ collider.~\cite{oddone}  At such a
machine, since one beam has higher energy than the other, the center of
mass is moving in the laboratory.  Thus, the $\Upsilon$(4S) and its
decay $B$'s are moving in the laboratory.  In this environment, the
observation of isolated secondary vertices is again a most useful tool.
Two major facilities of this sort have just turned on successfully, and
the experiment at each facility has reported its first physics results 
at the ICHEP 2000 meeting in Osaka, Japan.  The experiment is called
BaBar at the PEP-II machine at SLAC, and BELLE at the KEK-B machine in
Japan.

At fixed-target experiments, where the $b$ was first discovered, the
available energy is not as well controlled as at $e^+e^-$ colliders, but
the typical energies also produce $b$'s nearly at rest in the center of
mass.  However, because of the beam momentum, the $B$'s travel with
several tens of GeV/$c$ in the laboratory.  Several experiments have
observed bottom particles, especially the $\Upsilon$ and $B$-decays to
$J/\psi$'s.  E288, E605, E653, E771, and E789 at Fermilab are described
briefly in a commemorative book available on the web~\cite{ftceleb}.  
There are also HERA-B at DESY and BEATRICE (WA92) at the CERN SPS.   
Only HERA-B among all these is still in the data taking mode.

Much higher energy $B$'s and b-baryons are seen at high-energy hadron
collider experiments --  e.g., CDF, and DZero already, and at LHCb, BTeV
in the future.  Here the bottom particles have significant lifetimes in
the laboratory, making their observation easier -- except that the b
quarks often turn into full-scale jets of many particles, not just the
pristine $B$'s seen at lower energies.  Symmetric $e^+e^-$ colliders
running at the $Z^o$ mass and above have similar access to $b$ physics,
again with $b$'s dominantly in jets of particles.  We have seen
interesting results from the SLD experiment at SLAC and four LEP
experiments at CERN: ALEPH, DELPHI, L3, and OPAL.

Across this range of experiments, there are a few special techniques
which come into play.  At the $e^+e^-$ colliders running at the
$\Upsilon(4S)$, backgrounds are reduced and kinematic parameter
resolution improves using the ``beam constrained mass'' -- as was the
case for charm from the $\psi$'' experiments at $e^+e^-$ colliders.  In
addition, at fixed-target experiments and within jets at higher
energies, leptons with high $p_t$ play a role, and visible lifetimes for
open bottom mesons and baryons are again crucial.
  
In looking for mixing of neutral $B$ mesons in hadronic interactions, it
is again necessary to know the nature of the $B$ at birth, as well as
its nature when it decays.  In the case of the $e^+e^-$ colliders, the
difference in the nature of the two $B$'s as they decay is needed.  
This sort of knowledge is obtained by observing one $B$ as it decays,
and using information either about the other $B$ in the event or about
the produciton of the first $B$.  Using partial information about the 
other $B$ is called ``tagging.''  Tagging usually involves incomplete
knowledge, in order to have as many tagged events as possible.  Thus,
experimenters examine the net charge, weighted by momentum, or
observed decay leptons from the other $B$ or $b$-jet.  For information
about the fully reconstructed $B$, one can use information on its
production angle when there is a strong asymmetry, as when highly
polarized electron or positron beams are used at $e^+e^-$ colliders
operating at the $Z^o$ with its weak decay asymmetry --
so far only at the SLC facility at SLAC.  At hadron colliders one can
also examine other particles produced in the jet with the $B$.  The most
information is available when examining the single other particle
nearest in phase space to the $B$.  Detailed information about such
production correlations is then needed to fully understand the tagging
rate and backgrounds.

\subsection*{Physics Issues for Bottom Quarks}

The bottom quark is part of a standard (left-handed) quark isospin
doublet ($b$, $t$), although there were many early attempts of think of
it as a singlet.  The dominant $b$ decay is to $c$ quarks, via emission
of a virtual $W$, a $3^{rd}$ to $2^{nd}$ generation transition.  This
coupling is much less strong than the $s$ to $u$ transition, a $2^{nd}$ 
to $1^{st}$ generation transition.  The coupling of the $b$ to the $u$
quark, $3^{rd}$ to $1^{st}$ generation, is even weaker.  The GIM
mechanism breaks down since the $t$ quark is so massive.  The
massiveness of the $t$ quark also leads to very large
$B^o\overline{B}^o$ mixing.   This was observed very early via a
surprisingly large same-sign di-lepton signal,\cite{UA1samesign} though
it was not so widely accepted at first.  It was the first indication
that the top quark was so extraordinarily heavy.

Today, there are extensive efforts to measure the parameters of the $B$
meson system.  While we refer to this as being done to test the 
Standard Model, most hope that we will find evidence for physics beyond
the Standard Model -- that is, evidence that there is no single set of
parameters within the Standard Model that can explain all observations.
Thus, it is important to observe each parameter in more than a single
way.  This is often called overdetermining the CKM matrix.  Experiments
are focusing on:\\

\hspace{3cm} CP violation \\

\hspace{3cm} Oscillations in neutral states, both $B_d$ and $B_s$   \\

\hspace{3cm} Rare and nominally-forbidden decays \\

There are still issues in the details of decay dynamics, since
there are discrepancies between expectations and some lifetime
measurements.  An additional wrinkle in the picture is the relationship
between the semileptonic decay rate and the number of charm particles
observed in $b$ decay.  The semileptonic decay rate is $1$ to $2 \%$
below the rate (more like $12 \%$) expected from theory using
standard CKM matrix elements.  Considering $b \rightarrow W^-c$
and $W^-u$, with $W^- \rightarrow \overline{u} d$, $\overline{c} s$,
$e^- \overline{\nu}_e$, $\mu^- \overline{\nu}_{\mu}$, and $\tau^-
\overline{\nu}_{\tau}$, the semileptonic branching ratio should be
$\sim$ 1/9, given 3 colors for the quarks.  Phase space for $c$ and $t$
quarks and final state interactions for $W \rightarrow$ quarks modify
this somewhat.

In addition to the electroweak theory which is the main part of the
Standard Model being tested above, there are also issues for QCD.  
The flavor independence of the strong coupling constant has been 
established by looking at the rate of gluon emission in $e^+e^-$
interactions involving the light, the charm, and the bottom quark.
One might expect that the bottom quark is heavy enough that NLO
perturbative QCD would suffice to explain $b$ quark production.
However, again there are mysteries to be unraveled here.  The
importance of the hadronization process for tagging has been mentioned
already.  

\subsection*{Bottom Production - Cross Sections}

Given that the $b$ mass is much larger than $\Lambda_{QCD}$, the
expectation is that NLO perturbation theory will reliably predict 
$\sigma_{b\overline{b}}$.  At the Tevatron Collider, where the high
transverse momentum of the observed $B$'s should also help, the
measurements give $\sigma_{b\overline{b}}$ too large by a factor
at least two.\cite{CDFb,D0b}

\begin{figure}
\centerline{ \epsfxsize=5.5in \epsffile{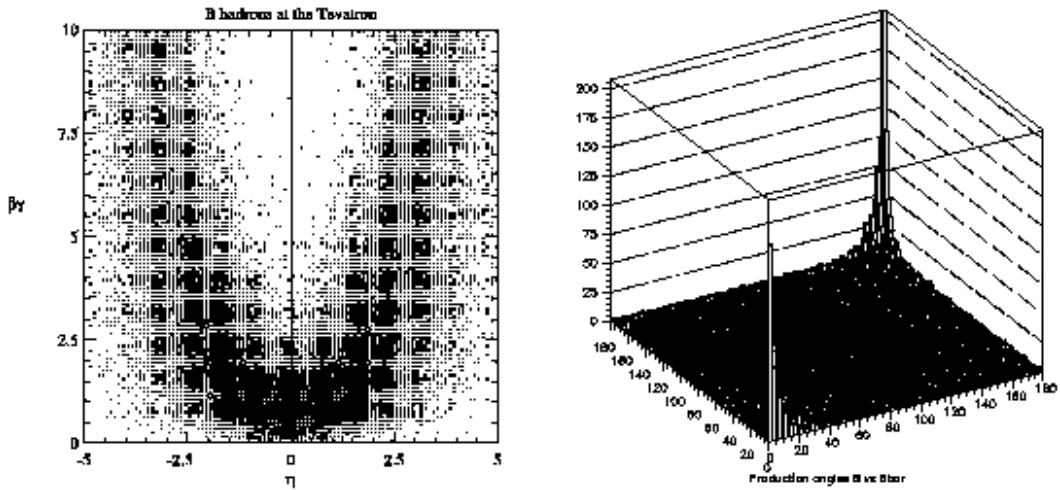}}
\caption{Distributions of $b$ and $\overline{b}$ produced in
interactions at the Tevatron Collider. On the left is the time dilation
factor $\gamma \beta$ vs the pseudorapidity, on the right is the
angular correlation of ${\it b}$ and ${\overline {\it b}}$ quarks.}
\label{bb}
\end{figure}

The CDF and DZero measurements at the Tevatron are dominantly in the
central region of phase space.  This will also be true for the LHC
experiments ATLAS and CMS.  However there are two new experiments which
will explore the forward regions, BTeV at the Tevatron and LHC-b at
the LHC.  These latter experiments will take advantage of the higher
momentum and longer laboratory lifetimes of bottom particles in their
kinematic region.  See the left side drawing of Fig. \ref{bb}.  Also
shown in Fig. \ref{bb} is the strong correlation in produciton of the
$b$ and $\overline{b}$ in the forward and backward directions predicted
by the PYTHIA simulation model.  This appears as the strong peaking in
the number of events shown on the figure on the right side of
Fig. \ref{bb}.  BTeV and LHC-b should be able to take advantage of this
correlation in tagging the $B$'s which they will observe.
   
\subsection*{Bottom Production - Hadronization Effects}

Bottom quarks turn into $B$ mesons which decay to charm.  The charm
distributions depend on the fragmentation (hadronization) functions for
the $b$ quarks, the harder the $B$ meson, the harder the charm. 
Examples are the \\

\noindent Andersson function:      
\begin{equation}
   \phi(z) \sim  z^{-1} (1-z)^a  e^{-b m_t^2/z}
\end{equation}
\noindent and the Peterson function:      
\begin{equation} 
   \phi(z) \sim  1 / [ z [(1 - (1/z) - \epsilon_P/(1-z)]^2]
\end{equation}

The CLEO Collaboration has looked at $D_s$ decays of $B$ mesons as a 
particularly clean way to determine the best parameters and relative
merits of these two functions.~\cite{briere}  These distributions are
important in understanding experiment apparatus acceptance and in
determining backgrounds for top quark and other heavy objects.  Yet, we
may ask whether the fragmentation functions from $e^+e^-$ are relevant
in hadronic environments?\cite{CDFfrag}  If they may be so at high $p_t$
where approximate isolation may be achieved, how low can one go in $p_t$
before the color-field environment is so different that the $e^+e^-$
measurements are not applicable.

\subsection*{Bottom-Particle Lifetimes}


%

The original expectations for bottom particles was that their lifetimes
would be short, even with a new quantum number to conserve, because of
the large phase space and multitude of decay modes available.   However,
it turned out that the CKM-matrix parameters were small, thus leading to
long lives.  As with charm, the expectation was for equality of
lifetimes -- all $\sim$ 1.5 picosecond.  However, as seen in Table
\ref{bparts}, there are significant differences.  The shortness of the
$\Lambda_b$ lifetime is particularly difficult to explain. 

\subsection*{Neutral Bottom-Particle Mixing}

Both $B_d$ and $B_s$ particles are expected to have observable
mixing.  However, the rate of $B_s$ oscillation will be very much more
rapid, and correspondingly harder to measure.  $B_d$ mixing has been
observed by several experiments, in a number of ways -- even
directly.  No $B_s$ mixing has yet been seen.  This mixing is expected
to lead to CP violation due to the phase in the CKM matrix (i.e., a
non-zero value of $\rho$ in Eq. \ref{ckmmatrix}).

\subsection*{Additional Bottom Decay Issues}

The very hot topic in bottom particle decays is CP violation in $B$
meson decay.  The idea is to test the Standard Model prediction for
the violation -- in the hopes of finding a disagreement which would
provide a clue to physics beyond the Standard Model, just as that is the
hope in studies of charm decay.  We anticipate that very heavy particles
may make virtual contributions to the box diagrams in a measureable
way.  Non-Standard Model ``penguin'' decay diagrams also are open to
virtual particles; e.g., heavy $W$ and charged Higgs in place of the
traditional $W$ in loops.  See the lectures at this workshop by Gabriel
Lopez, ``Violaci\'on de CP en Mesones B,'' for more on this subject.

Among the things which complicate experimental work in bottom decays is
the large number of decay channels available for each bottom
particle.  This leads to very small BR's (typically very much less than
1 \%) and the difficulty of collecting large samples of a given decay
mode.  Yet, this is exactly what is needed  for high precision
measurements and searches; e.g., searches for small CP asymmetries.  It
is usually the case that the largest such asymmetries  are anticipated
for the smaller branching-ratio decay modes.  Furthermore, only a
fraction of the decay channels have actually been observed.

\section*{Observations and Physics of Top Quarks}                   

Again, for the heaviest of quarks, the top quark, I will review its
first observation, the techniques used, the measurements made, and make
a rapid tour through the physics issues related directly to top quarks.

\subsection*{First Observations of Top Quarks}

The discovery of top quarks required seeing interactions with the
characteristics expected, and at a rate greater than that which could be
expected from other processes.  The signal for top quarks so far does
not stand out quite as clearly as the signals for charm and bottom
quarks did.  Given their very high mass, top quarks could only be
discovered in proton-antiproton collisions at the 1.8 TeV center-of-mass
energy available at Fermilab's Tevatron.  Furthermore, the signals
appear near the end of the spectra where the rates are low.  The
cross section turns out to be about $5 pb^-1$.  This corresponds to one
such event produced in about $2 \times 10^{10}$ events.

Observation of the top quarks has depended on the high mass of the top
quark, and the high mass and $p_t$ of its decay products.  Fortunately,
the decay is completely dominated by a single mode:

\begin{center}
$t \rightarrow W b$
\end{center}
Three topologies result from $ t \overline{t}$ production and decay: 
6 jets, 4 jets plus a lepton, and 2 jets plus 2 leptons:  \\

\hspace{3cm} $ t \overline{t} \rightarrow [W^- \overline{b}]
\hspace{0.2cm} [W^+ b]  
\rightarrow [(q \overline{q}')  \hspace{0.2cm} b]  
\hspace{0.3cm} [q''\overline{q}''')  \hspace{0.2cm} \overline{b}] $ \\

\hspace{3cm} $ t \overline{t} \rightarrow [W^- \overline{b}]
\hspace{0.2cm}[W^+ b] 
\rightarrow [(\ell \overline{\nu}_{\ell})  \hspace{0.2cm} b]
\hspace{0.3cm} [(q \overline{q}') \hspace{0.2cm} b] $ \\

$\hspace{6.2cm}$ and $[(q \overline{q}') \hspace{0.2cm} \overline{b}]
\hspace{0.3cm} [(\overline{\ell} \nu_{{\overline{\ell}}}) \hspace{0.2cm}
\overline{b}] $\\

\hspace{3cm} $ t \overline{t} \rightarrow [W^- \overline{b}]
\hspace{0.2cm}[W^+ b]
\rightarrow [(\ell \overline{\nu}_{\ell})  \hspace{0.2cm} b ]
\hspace{0.3cm} [(\overline{\ell}'\nu_{\overline{\ell}'})  \hspace{0.2cm}
\overline{b}] $ \\

In order to optimize the observation above background, the experiments
which made the observations, CDF and DZero, focused on central 
kinematic region.  Most other processes are peaked forward-backward; and
the high mass decay daughters benefit from the Jacobian enhancement 
perpendicular to the incident particles.  Only events with very high
$p_t$ jets and leptons; and missing, high transverse energy $E_t$ from
neutrinos were examined.  Furthermore, $b$-quark jet tagging via leptons
and secondary vertices was very helpful.  The largest backgrounds came
from $W$ plus low-mass $q$ jets.  So, the techniques used for bottom
quark physics are directly relevant here for top quarks as well.  
Finally, kinematic constraints on the $t$-quarks and $W^+$ and $W^-$
candidates were used.  Though the $t$ quark mass was unknown, both $t$'s
must have same mass; and $W$ mass was known.

\subsection*{Top Quark Production Cross Section}

Table \ref{t-sigma} gives the observed cross section results from both
CDF and DZero.  The events listed in the table come from approximately
100 $pb^{-1}$ of data, meaning that about 500 $t\overline{t}$ events
were produced in each experiment.  Thus, CDF and DZero saw about 10 \%
of the top events produced, even after their final selection criteria
were applied.  That is quite efficient for such rare events.  For those
signatures with the most observed events, the backgrounds are the
greatest.  Nevertheless, the various data sets are consistent with each 
other, and consistently above the expected backgrounds.

\begin{table}[t!]
\caption{Top Quark Production Cross Section Measurements.}
\begin{tabular}{|c|c|c|c|c|c|}
\hline
$\sigma_{t\overline{t}}$& Source&   Ref.   &      Method        & Total
&
Background\\
 (pb)              &       &          &                    & Events &
Events \\
\hline\hline
$4.1 \pm 2.0 $     & DZero & \cite{D0topX} & lepton plus jets      & 19&
  $ 8.7 \pm 1.7 $ \\
$8.2 \pm 3.5 $     & DZero & \cite{D0topX} & lepton plus jets/$\mu$& 11&
  $ 2.4 \pm 0.5 $ \\
$6.3 \pm 3.3 $     & DZero & \cite{D0topX} & dileptons and $e\nu$  &  9&
  $ 2.6 \pm 0.6 $ \\
$5.5 \pm 1.8 $     & DZero & \cite{D0topX} &  DZero combined	   & 39&
  $13.7 \pm 2.2 $ \\
\hline
$6.7^{+2.0}_{-1.7}$  &  CDF  & \cite{CDFtopX1} & lepton plus jets  & 34&
  $ 9.2 \pm 1.5 $ \\
                   &       &                 &                     & 40&
  $22.6 \pm 2.8 $ \\
$8.2^{+4.4}_{-3.4}$&  CDF  & \cite{CDFtopX2} & dileptons           &  9&
  $ 2.4 \pm 0.5 $ \\
$10.1^{+4.5}_{-3.6}$& CDF  & \cite{CDFtopX3} & all jets , $>$0 tags&187&
  $ 142 \pm 12  $ \\
                   &       &          &  all jets , $>$1 tag       &157&
  $ 120 \pm 18  $ \\
$7.6^{+1.8}_{-1.5}$&  CDF  & \cite{CDFtopX1} & CDF combined        &   &
                  \\
\hline\hline
$  4.7 - 5.8    $  & Theory& \cite{THtopX}&for $m_t$ 173-175 GeV/$c^2$&&
                  \\
\hline
\end{tabular}
\label{t-sigma}
\end{table}

\subsection*{Top Quark Mass Measurement}

As suggested in the discussion of the cross section, kinematic
fits to the events can leave the value of the $t$-quark mass free, only
demanding that both the $t$ and $\overline{t}$ have the same mass.
Table \ref{t-mass} shows the results of the fits of essentially the same
events as were used for the cross section measurements.  Again,
agreement of the various sets and between the two experiments gives
confidence in the basic result.

\begin{table}[t!]
\caption{Top Quark Mass Measurements.}
\centering
\begin{tabular}{|c|c|c|c|c|c|}
\hline
$ m_t$ (GeV/$c^2$)      &Source&   Ref.         &      Method
& 
Total & Background       \\   
 (pb)                   &      &                &                     & 
Events&           Events \\
\hline\hline
$173.3 \pm 5.6 \pm 5.5 $&DZero & \cite{D0topM1} & lepton plus jets    &
   76 &$53.2^{+11}_{-9} $\\
                        &      &                &                     & 
      &$48.2^{+11}_{-9}$ \\
$168.4 \pm 12.3 \pm 3.6$&DZero & \cite{D0topM2} & dileptons           &
    6 &                  \\
$172.1 \pm 5.2 \pm 4.9 $&DZero & \cite{D0topM1} & DZero combined      &
      &                  \\
\hline
$175.9 \pm 4.8 \pm 4.9 $&  CDF & \cite{CDFtopM1}& lepton plus jets    &
   76 &  $31 \pm 8 $     \\
$161 \pm 17 \pm 10     $&  CDF & \cite{CDFtopM2}& dileptons           &
    9 &  $ 2.4 \pm 0.5 $ \\
$186 \pm 10 \pm 12 $    &  CDF & \cite{CDFtopM3}&all jets , $>$0 tags &
  187 &  $ 142 \pm 12  $ \\
                        &      &                & all jets, $>$ 1 tag &
  157 &  $ 120 \pm 18  $ \\
\hline
$173.8 \pm 3.5 \pm 3.9 $&  PDG & \cite{PDGtopM} & PDG Average         &
      &                  \\
\hline
\end{tabular}
\label{t-mass}
\end{table}

\subsection*{Relation Between Mass and Cross Section and Other Tests}

Given that the top quark mass is above the $Wb$ threshold, the decay
width should be about 1 GeV, corresponding to a lifetime of about 
$10^{-23}$ seconds.  Thus, the $t$ decays before top-hadrons or onia
form.  This is why the decay is totally dominated by the $W$ $b$ decay
mode.  The production cross section is calculated in NLO perturbative
QCD.  The perturbative calculation is expected to be very reliable.  It
does depend on the mass of the top quark, of course.  Any inconsistency
of the production rate and measured mass would cast doubt on the
discovery.  No discrepancy is observed, as seen in Table \ref{t-sigma}.

On the other hand, discrepancies might indicate exotic production
channels, as could an unexpected distribution of the measured 
$t\overline{t}$ mass distribution and the distribution of the transverse
momentum of the $t$. 
Angular distributions are sensitive to the presumed V-A coupling and
the relative coupling of transverse and longitudinal $W$'s to
$t$.  The fraction of decays to transversely and longitudinally
polarized $W$'s for 175 GeV/$c^2$ $t$ is 30 \%.  A discrepancy would
challenge the Higgs mechanism of spontaneous symmetry breaking.
Finally, one can also imagine seeing clues to new physics via rare or
forbidden decay rates.  Finally, the top could be the source of
observation of non-Standard Model particles.   For example, if the   
Higgs particle were light enough, the top quark could decay emitting a
Higgs particle.  One could also be sensitive to decays to light enough
SUSY particles.

\section*{The Next Heavy Particles, Their Observation and Nature}

In the search for heavier objects, heavy quarks will play a critical
role as tags for the discovery of those heavier objects, and as a
key to identifying what they are.  We have seen how this was the case in
the top discovery, where heavy quarks in jets separated top events from
others among the highest $p_t$ interactions.

\subsection*{Predictions for the Higgs Particle}

The Higgs mechanism for electroweak symmetry breaking manifests
itself in a particle whose coupling to another particle is proportional
to the other particle's mass.  Thus, Standard Model predictions of the
Higgs mass can be made via virtual loop corrections for mass of the top
quark.  Note that the dependence of the $W$ mass on the top mass is
linear, while the dependence is logarithmic on the Higgs mass.  Today,
there is a lot of focus on this prediction, as the competition between
Fermilab and CERN to discover the Higgs is very intense.  The Higgs
branching ratio to other particles depends, also, on the Higgs mass --
due to threshold and phase space effects.  So, identifying a signal at a
given mass as a Higgs particle depends on measuring the relative
branching ratios to as many of the heavy quarks, $WW$, $WW^*$, $ZZ$, and
$ZZ^*$ states as are kinematically allowed.  The asterisks in decays to
$WW^*$ and $ZZ^*$ refer to virtual, intermediate decay particles, 
$W^*$ and $Z^*$ whose decay can be to kinematically allowed final 
states.  Many physicists expect that direct observation of a Higgs
particle will be the next breakthrough discovery.  On the other hand,
surprises can happen.  

\subsection*{Only Three Generations of Quarks?}

Are the three generations of quark pairs truly the complete story?  The
limits on additional light neutrinos from $Z^o$ decay do not necessarily
prove this, since there could be a very heavy neutrino for a fourth
generation.  There are some limits on this from virtual effects in
rare decays of heavy quarks.

\subsection*{Characterizing the Squarks and Higgs of SUSY}

There is not enough time in these lectures to give an overview of
SUSY.  The topic has been well covered in the lectures of Cupatitzio
Ram\'{\i}rez and Carlos Wagner.  Let me only note that the host of
predicted SUSY particles will place a premium on the detailed
characterization of any candidate observations, as well as filling out
the mass-spectrum of particles discovered.  Thus, the heavy squark
decays to the corresponding quark, as well as SUSY Higgs couplings
are critical to the understanding of what will have been seen.

\section*{Concluding Comments}                   

I would like to recommend some reading for more detail than I could
give here, in particular: a review article by Harry Lipkin~\cite{lipkin}
for the state of thinking before the November Revolution of 1974, Ed
Thorndike's summary of bottom quark physics~\cite{thorndike}, and the
PDG
summary by Michelangelo Mangano and Thomas Trippe~\cite{THtop} on top
quarks.

We have seen that heavy quarks have been both interesting in their own
right, and useful tools for understanding other things -- from 
light-meson resonances to the high-mass extensions of the current
Standard Model.  I want to note again the importance played by applying
new technology matched to physics opportunities.  The precision tracking
made possible by microelectronic advances would have been less
interesting except for the few mm lifetimes of the charm and bottom
quarks in the laboratory.  The fantastic increase in computing and
storage efficiency made possible the treatment of large amounts of data
as required by the low cross sections for heavy quark production.
Finally, we should also be a little humble amid all the excitement of
the anticipated discoveries of the next heavy objects.  In my brief
review, we have seen many of the results fly in the face of ``common
knowledge'' and general expectations.

Let me close by reiterating the wealth of rather recent progress in
understanding the quark nature of matter, and the enjoyment that has
come from being part of this, both from the physics and from technical
efforts.  I also want to thank our hosts and organizers for a most
stimulating and enjoyable workshop.

\end{document}